# Cloud Computing and Grid Computing 360-Degree Compared

[1,2,3]Ian Foster, [4]Yong Zhao, [1]Ioan Raicu, [5]Shiyong Lu

*foster@mcs.anl.gov, yozha@microsoft.com, iraicu@cs.uchicago.edu, shiyong@wayne.edu*

[1] Department of Computer Science, University of Chicago, Chicago, IL, USA
[2] Computation Institute, University of Chicago, Chicago, IL, USA
[3] Math & Computer Science Division, Argonne National Laboratory, Argonne, IL, USA
[4] Microsoft Corporation, Redmond, WA, USA
[5] Department of Computer Science, Wayne State University, Detroit, MI, USA

*Abstract*– Cloud Computing has become another buzzword after Web 2.0. However, there are dozens of different definitions for Cloud Computing and there seems to be no consensus on what a Cloud is. On the other hand, Cloud Computing is not a completely new concept; it has intricate connection to the relatively new but thirteen-year established Grid Computing paradigm, and other relevant technologies such as utility computing, cluster computing, and distributed systems in general. This paper strives to compare and contrast Cloud Computing with Grid Computing from various angles and give insights into the essential characteristics of both.

## 1 100-Mile Overview

Cloud Computing is hinting at a future in which we won't compute on local computers, but on centralized facilities operated by third-party compute and storage utilities. We sure won't miss the shrink-wrapped software to unwrap and install. Needless to say, this is not a new idea. In fact, back in 1961, computing pioneer John McCarthy predicted that "computation may someday be organized as a public utility"—and went on to speculate how this might occur.

In the mid 1990s, the term Grid was coined to describe technologies that would allow consumers to obtain computing power on demand. Ian Foster and others posited that by standardizing the protocols used to request computing power, we could spur the creation of a Computing Grid, analogous in form and utility to the electric power grid. Researchers subsequently developed these ideas in many exciting ways, producing for example large-scale federated systems (TeraGrid, Open Science Grid, caBIG, EGEE, Earth System Grid) that provide not just computing power, but also data and software, on demand. Standards organizations (e.g., OGF, OASIS) defined relevant standards. More prosaically, the term was also co-opted by industry as a marketing term for clusters. But no viable commercial Grid Computing providers emerged, at least not until recently.

So is "Cloud Computing" just a new name for Grid? In information technology, where technology scales by an order of magnitude, and in the process reinvents itself, every five years, there is no straightforward answer to such questions.

*Yes*: the vision is the same—to reduce the cost of computing, increase reliability, and increase flexibility by transforming computers from something that we buy and operate ourselves to something that is operated by a third party.

*But no*: things are different now than they were 10 years ago. We have a new need to analyze massive data, thus motivating greatly increased demand for computing. Having realized the benefits of moving from mainframes to commodity clusters, we find that those clusters are quite expensive to operate. We have low-cost virtualization. And, above all, we have multiple billions of dollars being spent by the likes of Amazon, Google, and Microsoft to create real commercial large-scale systems containing hundreds of thousands of computers. The prospect of needing only a credit card to get on-demand access to 100,000+ computers in tens of data centers distributed throughout the world—resources that be applied to problems with massive, potentially distributed data, is exciting! So we are operating at a different scale, and operating at these new, more massive scales can demand fundamentally different approaches to tackling problems. It also enables—indeed is often only applicable to—entirely new problems.

*Nevertheless, yes*: the problems are mostly the same in Clouds and Grids. There is a common need to be able to manage large facilities; to define methods by which consumers discover, request, and use resources provided by the central facilities; and to implement the often highly parallel computations that execute on those resources. Details differ, but the two communities are struggling with many of the same issues.

### 1.1 Defining Cloud Computing

There is little consensus on how to define the Cloud [49]. We add yet another definition to the already saturated list of definitions for Cloud Computing:

> *A large-scale distributed computing paradigm that is driven by economies of scale, in which a pool of abstracted, virtualized, dynamically-scalable, managed computing power, storage, platforms, and services are delivered on demand to external customers over the Internet.*

There are a few key points in this definition. First, Cloud Computing is a specialized distributed computing paradigm; it differs from traditional ones in that 1) it is massively scalable, 2) can be encapsulated as an abstract entity that delivers different levels of services to customers outside the Cloud, 3) it is driven by economies of scale [44], and 4) the services can be dynamically configured (via virtualization or other approaches) and delivered on demand.

Governments, research institutes, and industry leaders are rushing to adopt Cloud Computing to solve their ever-increasing computing and storage problems arising in the Internet Age. There are three main factors contributing to the surge and interests in Cloud Computing: 1) rapid decrease in hardware cost and increase in computing power and storage capacity, and the advent of multi-core architecture and modern supercomputers consisting of hundreds of thousands of cores;

2) the exponentially growing data size in scientific instrumentation/simulation and Internet publishing and archiving; and 3) the wide-spread adoption of Services Computing and Web 2.0 applications.

## 1.2 Clouds, Grids, and Distributed Systems

Many discerning readers will immediately notice that our definition of Cloud Computing overlaps with many existing technologies, such as Grid Computing, Utility Computing, Services Computing, and distributed computing in general. We argue that Cloud Computing not only overlaps with Grid Computing, it is indeed evolved out of Grid Computing and relies on Grid Computing as its backbone and infrastructure support. The evolution has been a result of a shift in focus from an infrastructure that delivers storage and compute resources (such is the case in Grids) to one that is economy based aiming to deliver more abstract resources and services (such is the case in Clouds). As for Utility Computing, it is not a new paradigm of computing infrastructure; rather, it is a business model in which computing resources, such as computation and storage, are packaged as metered services similar to a physical public utility, such as electricity and public switched telephone network. Utility computing is typically implemented using other computing infrastructure (e.g. Grids) with additional accounting and monitoring services. A Cloud infrastructure can be utilized internally by a company or exposed to the public as utility computing.

See Figure 1 for an overview of the relationship between Clouds and other domains that it overlaps with. Web 2.0 covers almost the whole spectrum of service-oriented applications, where Cloud Computing lies at the large-scale side. Supercomputing and Cluster Computing have been more focused on traditional non-service applications. Grid Computing overlaps with all these fields where it is generally considered of lesser scale than supercomputers and Clouds.

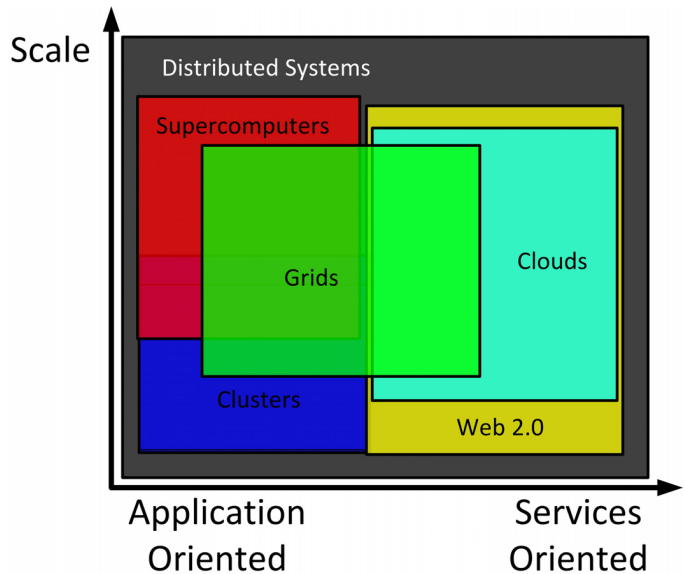

**Figure 1: Grids and Clouds Overview**

Grid Computing aims to "enable resource sharing and coordinated problem solving in dynamic, multi-institutional virtual organizations" [18][20]. There are also a few key features to this definition: First of all, Grids provide a distributed computing paradigm or infrastructure that spans across multiple virtual organizations (VO) where each VO can consist of either physically distributed institutions or logically related projects/groups. The goal of such a paradigm is to enable federated resource sharing in dynamic, distributed environments. The approach taken by the *de facto* standard implementation – The Globus Toolkit [16][23], is to build a uniform computing environment from diverse resources by defining standard network protocols and providing middleware to mediate access to a wide range of heterogeneous resources. Globus addresses various issues such as security, resource discovery, resource provisioning and management, job scheduling, monitoring, and data management.

Half a decade ago, Ian Foster gave a three point checklist [19] to help define what is, and what is not a *Grid*: 1) coordinates resources that are not subject to centralized control, 2) uses standard, open, general-purpose protocols and interfaces, and 3) delivers non-trivial qualities of service. Although point 3 holds true for Cloud Computing, neither point 1 nor point 2 is clear that it is the case for today's Clouds. The vision for Clouds and Grids are similar, details and technologies used may differ, but the two communities are struggling with many of the same issues. This paper strives to compare and contrast Cloud Computing with Grid Computing from various angles and give insights into the essential characteristics of both, with the hope to paint a less cloudy picture of what Clouds are, what kind of applications can Clouds expect to support, and what challenges Clouds are likely to face in the coming years as they gain momentum and adoption. We hope this will help both communities gain deeper understanding of the goals, assumptions, status, and directions, and provide a more detailed view of both technologies to the general audience.

## 2 Comparing Grids and Clouds Side-by-Side

This section aims to compare Grids and Clouds across a wide variety of perspectives, from architecture, security model, business model, programming model, virtualization, data model, compute model, to provenance and applications. We also outline a number of challenges and opportunities that Grid Computing and Cloud Computing bring to researchers and the IT industry, most common to both, but some are specific to one or the other.

### 2.1 Business Model

Traditional business model for software has been a one-time payment for unlimited use (usually on 1 computer) of the software. In a cloud-based business model, a customer will pay the provider on a consumption basis, very much like the utility companies charge for basic utilities such as electricity, gas, and water, and the model relies on economies of scale in order to drive prices down for users and profits up for providers. Today, Amazon essentially provides a centralized Cloud consisting of Compute Cloud EC2 and Data Cloud S3. The former is charged based on per instance-hour consumed for each instance type and the later is charged by per GB-Month of storage used. In addition, data transfer is charged by TB / month data transfer, depending on the source and target of such transfer. The prospect of needing only a credit card to get on-demand access to 100,000+ processors in tens of data centers distributed throughout the world—resources that be applied to problems with massive, potentially distributed data, is exciting!

The business model for Grids (at least that found in academia or government labs) is project-oriented in which the users or community represented by that proposal have certain number of service units (i.e. CPU hours) they can spend. For example, the TeraGrid operates in this fashion, and requires increasingly complex proposals be written for increasing number of computational power. The TeraGrid has more than a dozen Grid sites, all hosted at various institutions around the country. What makes an institution want to join the TeraGrid? When an institution joins the TeraGrid with a set of resources, it knows that others in the community can now use these resources across the country. It also acknowledges the fact that it gains access to a dozen other Grid sites. This same model has worked rather well for many Grids around the globe, giving institutions incentives to join various Grids for access to additional resources for all the users from the corresponding institution.

There are also endeavors to build a Grid economy for a global Grid infrastructure that supports the trading, negotiation, provisioning, and allocation of resources based on the levels of services provided, risk and cost, and users' preferences; so far, resource exchange (e.g. trade storage for compute cycles), auctions, game theory based resource coordination, virtual currencies, resource brokers and intermediaries, and various other economic models have been proposed and applied in practice [8].

## 2.2 Architecture

Grids started off in the mid-90s to address large-scale computation problems using a network of resource-sharing commodity machines that deliver the computation power affordable only by supercomputers and large dedicated clusters at that time. The major motivation was that these high performance computing resources were expensive and hard to get access to, so the starting point was to use federated resources that could comprise compute, storage and network resources from multiple geographically distributed institutions, and such resources are generally heterogeneous and dynamic. Grids focused on integrating existing resources with their hardware, operating systems, local resource management, and security infrastructure.

In order to support the creation of the so called "Virtual Organizations"—a logical entity within which distributed resources can be discovered and shared as if they were from the same organization, Grids define and provide a set of standard protocols, middleware, toolkits, and services built on top of these protocols. Interoperability and security are the primary concerns for the Grid infrastructure as resources may come from different administrative domains, which have both global and local resource usage policies, different hardware and software configurations and platforms, and vary in availability and capacity.

Grids provide protocols and services at five different layers as identified in the Grid protocol architecture (see Figure 2). At the *fabric layer*, Grids provide access to different resource types such as compute, storage and network resource, code repository, etc. Grids usually rely on existing fabric components, for instance, local resource managers (i.e. PBS [5], Condor [48], etc). General-purpose components such as GARA (general architecture for advanced reservation) [17], and specialized resource management services such as Falkon [40] (although strictly speaking, Falkon also provides services beyond the fabric layer).

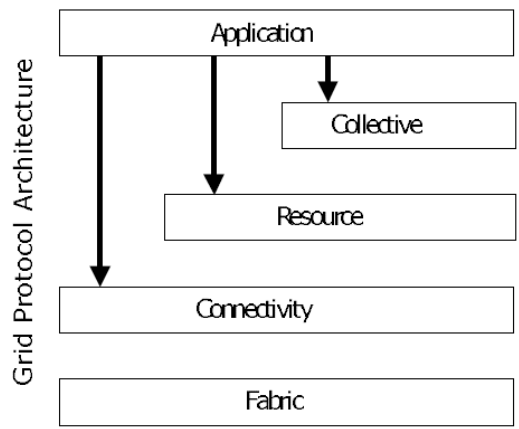

**Figure 2: Grid Protocol Architecture**

The *connectivity layer* defines core communication and authentication protocols for easy and secure network transactions. The GSI (Grid Security Infrastructure) [27] protocol underlies every Grid transaction.

The *resource layer* defines protocols for the publication, discovery, negotiation, monitoring, accounting and payment of sharing operations on individual resources. The GRAM (Grid Resource Access and Management) [16] protocol is used for allocation of computational resources and for monitoring and control of computation on those resources, and GridFTP [2] for data access and high-speed data transfer.

The *collective layer* captures interactions across collections of resources, directory services such as MDS (Monitoring and Discovery Service) [43] allows for the monitoring and discovery of VO resources, Condor-G [24] and Nimrod-G [7] are examples of co-allocating, scheduling and brokering services, and MPICH [32] for Grid enabled programming systems, and CAS (community authorization service) [21] for global resource policies.

The *application layer* comprises whatever user applications built on top of the above protocols and APIs and operate in VO environments. Two examples are Grid workflow systems, and Grid portals (i.e. QuarkNet e-learning environment [52], National Virtual Observatory (http://www.us-vo.org), TeraGrid Science gateway (http://www.teragrid.org)).

Clouds are developed to address Internet-scale computing problems where some assumptions are different from those of the Grids. Clouds are usually referred to as a large pool of computing and/or storage resources, which can be accessed via standard protocols via an abstract interface. Clouds can be built on top of many existing protocols such as Web Services (WSDL, SOAP), and some advanced Web 2.0 technologies such as REST, RSS, AJAX, etc. In fact, behind the cover, it is possible for Clouds to be implemented over existing Grid technologies leveraging more than a decade of community efforts in standardization, security, resource management, and virtualization support.

There are also multiple versions of definition for Cloud architecture, we define a four-layer architecture for Cloud Computing in comparison to the Grid architecture, composed of 1) fabric, 2) unified resource, 3) platform, and 4) application Layers.

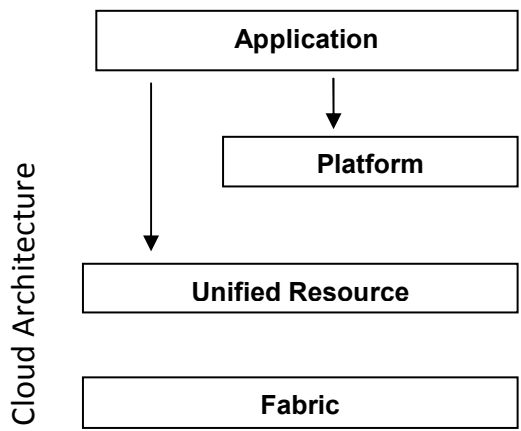

Figure 3: Cloud Architecture

The *fabric layer* contains the raw hardware level resources, such as compute resources, storage resources, and network resources. The *unified resource layer* contains resources that have been abstracted/encapsulated (usually by virtualization) so that they can be exposed to upper layer and end users as integrated resources, for instance, a virtual computer/cluster, a logical file system, a database system, etc. The *platform layer* adds on a collection of specialized tools, middleware and services on top of the unified resources to provide a development and/or deployment platform. For instance, a Web hosting environment, a scheduling service, etc. Finally, the *application layer* contains the applications that would run in the Clouds.

Clouds in general provide services at three different levels (*IaaS*, *PaaS*, and *Saas* [50]) as follows, although some providers can choose to expose services at more than one level. *Infrastructure as a Service (IaaS)* [50] provisions hardware, software, and equipments (mostly at the unified resource layer, but can also include part of the fabric layer) to deliver software application environments with a resource usage-based pricing model. Infrastructure can scale up and down dynamically based on application resource needs. Typical examples are Amazon EC2 (Elastic Cloud Computing) Service [3] and S3 (Simple Storage Service) [4] where compute and storage infrastructures are open to public access with a utility pricing model; Eucalyptus [15] is an open source Cloud implementation that provides a compatible interface to Amazon's EC2, and allows people to set up a Cloud infrastructure at premise and experiment prior to buying commercial services.

*Platform as a Service (PaaS)* [50] offers a high-level integrated environment to build, test, and deploy custom applications. Generally, developers will need to accept some restrictions on the type of software they can write in exchange for built-in application scalability. An example is Google's App Engine [28], which enables users to build Web applications on the same scalable systems that power Google applications.

*Software as a Service (SaaS)* [50] delivers special-purpose software that is remotely accessible by consumers through the Internet with a usage-based pricing model. Salesforce is an industry leader in providing online CRM (Customer Relationship Management) Services. Live Mesh from Microsoft allows files and folders to be shared and synchronized across multiple devices.

Although Clouds provide services at three different levels (*IaaS*, *PaaS*, and *Saas*), standards for interfaces to these different levels still remain to be defined. This leads to interoperability problems between today's Clouds, and there is little business incentives for Cloud providers to invest additional resources in defining and implementing new interfaces. As Clouds mature, and more sophisticated applications and services emerge that require the use of multiple Clouds, there will be growing incentives to adopt standard interfaces that facilitate interoperability in order to capture emerging and growing markets in a saturated Cloud market.

## 2.3 Resource Management

This section describes the resource management found in Grids and Clouds, covering topics such as the compute model, data model, virtualization, monitoring, and provenance. These topics are extremely important to understand the main challenges that both Grids and Clouds face today, and will have to overcome in the future.

**Compute Model:** Most Grids use a batch-scheduled compute model, in which a local resource manager (LRM), such as PBS, Condor, SGE manages the compute resources for a Grid site, and users submit batch jobs (via GRAM) to request some resources for some time. Many Grids have policies in place that enforce these batch jobs to identify the user and credentials under which the job will run for accounting and security purposes, the number of processors needed, and the duration of the allocation. For example, a job could say, stage in the input data from a URL to the local storage, run the application for 60 minutes on 100 processors, and stage out the results to some FTP server. The job would wait in the LRM's wait queue until the 100 processors were available for 60 minutes, at which point the 100 processors would be allocated and dedicated to the application for the duration of the job. Due to the expensive scheduling decisions, data staging in and out, and potentially long queue times, many Grids don't natively support interactive applications; although there are efforts in the Grid community to enable lower latencies to resources via multi-level scheduling, to allow applications with many short-running tasks to execute efficiently on Grids [40]. Cloud Computing compute model will likely look very different, with resources in the Cloud being shared by all users at the same time (in contrast to dedicated resources governed by a queuing system). This should allow latency sensitive applications to operate natively on Clouds, although ensuring a good enough level of QoS is being delivered to the end users will not be trivial, and will likely be one of the major challenges for Cloud Computing as the Clouds grow in scale, and number of users.

**Data Model:** While some people boldly predicate that future Internet Computing will be towards Cloud Computing centralized, in which storage, computing, and all kind of other resources will mainly be provisioned by the Cloud, we envision that the next-generation Internet Computing will take the triangle model shown in Figure 4: Internet Computing will be centralized around Data, Clouding Computing, as well as Client Computing. Cloud Computing and Client Computing will coexist and evolve hand in hand, while data management (mapping, partitioning, querying, movement, caching, replication, etc) will become more and more important for both Cloud Computing and Client Computing with the increase of data-intensive applications.

The critical role of Cloud Computing goes without saying, but the importance of Client Computing cannot be overlooked either for several reasons: 1) For security reasons, people might not be willing to run mission-critical applications on the Cloud and send sensitive data to the Cloud for processing and storage; 2) Users want to get their things done even when the Internet and Cloud are down or the network communication is slow; 3) With the advances of multi-core technology, the coming decade will bring the possibilities of having a desktop supercomputer with 100s to 1000s of hardware threads/cores. Furthermore, many end-users will have various hardware-driven end-functionalities, such as visualization and multimedia playback, which will typically run locally.

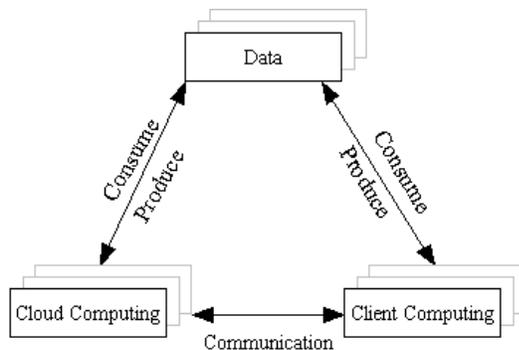

**Figure 4 The triangle model of next-generation Internet Computing.**

The importance of data has caught the attention of the Grid community for the past decade; Data Grids [10] have been specifically designed to tackle data intensive applications in Grid environments, with the concept of virtual data [22] playing a crucial role. Virtual data captures the relationship between data, programs and computations and prescribes various abstractions that a data grid can provide: location transparency where data can be requested without regard to data location, a distributed metadata catalog is engaged to keep track of the locations of each piece of data (along with its replicas) across grid sites, and privacy and access control are enforced; materialization transparency: data can be either re-computed on the fly or transferred upon request, depending on the availability of the data and the cost to re-compute. There is also representation transparency where data can be consumed and produced no matter what their actual physical formats and storage are, data are mapped into some abstract structural representation and manipulated in that way.

**Data Locality:** As CPU cycles become cheaper and data sets double in size every year, the main challenge for efficient scaling of applications is the location of the data relative to the available computational resources – moving the data repeatedly to distant CPUs is becoming the bottleneck. [46] There are large differences in IO speeds from local disk storage to wide area networks, which can drastically affect application performance. To achieve good scalability at Internet scales for Clouds, Grids, and their applications, data must be distributed over many computers, and computations must be steered towards the best place to execute in order to minimize the communication costs [46]. Google's MapReduce [13] system runs on top of the Google File System, within which data is loaded, partitioned into chunks, and each chunk replicated. Thus data processing is collocated with data storage: when a file needs to be processed, the job scheduler consults a storage metadata service to get the host node for each chunk, and then schedules a "map" process on that node, so that data locality is exploited efficiently. In Grids, data storage usually relies on a shared file systems (e.g. NFS, GPFS, PVFS, Luster), where data locality cannot be easily applied. One approach is to improve schedulers to be data-aware, and to be able to leverage data locality information when scheduling computational tasks; this approach has shown to improve job turn-around time significantly [41].

**Combining compute and data management:** Even more critical is the combination of the compute and data resource management, which leverages data locality in access patterns to minimize the amount of data movement and improve end-application performance and scalability. Attempting to address the storage and computational problems separately forces much data movement between computational and storage resources, which will not scale to tomorrow's peta-scale datasets and millions of processors, and will yield significant underutilization of the raw resources. It is important to schedule computational tasks close to the data, and to understand the costs of moving the work as opposed to moving the data. Data-aware schedulers and dispersing data close to processors is critical in achieving good scalability and performance. Finally, as the number of processor-cores is increasing (the largest supercomputers today have over 200K processors and Grids surpassing 100K processors), there is an ever-growing emphasis for support of high throughput computing with high sustainable dispatch and execution rates. We believe that data management architectures are important to ensure that the data management implementations scale to the required dataset sizes in the number of files, objects, and dataset disk space usage while at the same time, ensuring that data element information can be retrieved fast and efficiently. Grids have been making progress in combining compute and data management with data-aware schedulers [41], but we believe that Clouds will face significant challenges in handling data-intensive applications without serious efforts invested in harnessing the data locality of application access patterns. Although data-intensive applications may not be typical applications that Clouds deal with today, as the scales of Clouds grow, it may just be a matter of time for many Clouds.

**Virtualization:** Virtualization has become an indispensable ingredient for almost every Cloud, the most obvious reasons are for abstraction and encapsulation. Just like threads were introduced to provide users the "illusion" as if the computer were running all the threads simultaneously, and each thread were using all the available resources, Clouds need to run multiple (or even up to thousands or millions of) user applications, and all the applications appear to the users as if they were running simultaneously and could use all the available resources in the Cloud. Virtualization provides the necessary abstraction such that the underlying fabric (raw compute, storage, network resources) can be unified as a pool of resources and resource overlays (e.g. data storage services, Web hosting environments) can be built on top of them. Virtualization also enables each application to be encapsulated such that they can be configured, deployed, started, migrated, suspended, resumed, stopped, etc., and thus provides better security, manageability, and isolation.

There are also many other reasons that Clouds tend to adopt virtualization: 1) server and application consolidation, as multiple applications can be run on the same server, resources can be utilized more efficiently; 2) configurability, as the resource requirements for various applications could differ significantly, some require large storage, some compute, in order to dynamically configure and bundle (aggregate) resources for various needs, virtualization is necessary as this is not achievable at the hardware level; 3) increased application availability, virtualization allows quick recovery from unplanned outages, as virtual environments can be backed up and migrated with no interruption in service; 4) improved responsiveness: resource provisioning, monitoring and maintenance can be automated, and common resources can be cached and reused. All these features of virtualization provide the basis for Clouds to meet stringent SLA (Service Level Agreement) requirements in a business setting, which cannot be easily achieved with a non-virtualized environment in a cost-effective manner as systems would have to be over-provisioned to handle peak load and waste resources in idle periods. After all, a virtualization infrastructure can be just thought as a mapping from IT resources to business needs.

Grids do not rely on virtualization as much as Clouds do, but that might be more due to policy and having each individual organization maintain full control of their resources (i.e. by not virtualizing them). However, there are efforts in Grids to use virtualization as well, such as Nimbus [56] (previously known as the Virtual Workspace Service [26]), which provide the same abstraction and dynamic deployment capabilities. A virtual workspace is an execution environment that can be deployed dynamically and securely in the Grid. Nimbus provides two levels of guarantees: 1) quality of life: users get exactly the (software) environment they need, and 2) quality of service: provision and guarantee all the resources the workspace needs to function correctly (CPU, memory, disk, bandwidth, availability), allowing for dynamic renegotiation to reflect changing requirements and conditions. In addition, Nimbus can also provision a virtual cluster for Grid applications (e.g. a batch scheduler, or a workflow system), which is also dynamically configurable, a growing trend in Grid Computing.

It is also worth noting that virtualization – in the past – had significant performance losses for some applications, which has been one of the primary disadvantage of using virtualization. However, over the past few years, processor manufacturers such as AMD and Intel have been introducing hardware support for virtualization, which is helping narrow the performance gap between applications performance on virtualized resources as it compares with that on traditional operating systems without virtualization.

**Monitoring:** Another challenge that virtualization brings to Clouds is the potential difficulty in fine-control over the monitoring of resources. Although many Grids (such as TeraGrid) also enforce restrictions on what kind of sensors or long-running services a user can launch, Cloud monitoring is not as straightforward as in Grids, because Grids in general have a different trust model in which users via their identity delegation can access and browse resources at different Grid sites, and Grid resources are not highly abstracted and virtualized as in Clouds; for example, the Ganglia [25] distributed (and hierarchical) monitoring system can monitor a federation of clusters and Grids and has seen wide adoption in the Grid community. In a Cloud, different levels of services can be offered to an end user, the user is only exposed to a pre-defined API, and the lower level resources are opaque to the user (especially at the PaaS and SaaS level, although some providers may choose to expose monitoring information at these levels). The user does not have the liberty to deploy her own monitoring infrastructure, and the limited information returned to the user may not provide the necessary level of details for her to figure out what the resource status is. The same problems potentially exist for Cloud developers and administrators, as the abstract/unified resources usually go through virtualization and some other level of encapsulation, and tracking the issues down the software/hardware stack might be more difficult. Essentially monitoring in Clouds requires a fine balance of business application monitoring, enterprise server management, virtual machine monitoring, and hardware maintenance, and will be a significant challenge for Cloud Computing as it sees wider adoption and deployments. On the other hand, monitoring can be argued to be less important in Clouds, as users are interacting with a more abstract layer that is potentially more sophisticated; this abstract layer could respond to failures and quality of service (QoS) requirements automatically in a general-purpose way irrespective of application logic. In the near future, user-end monitoring might be a significant challenge for Clouds, but it will become less important as Clouds become more sophisticated and more self-maintained and self-healing.

**Provenance:** Provenance refers to the derivation history of a data product, including all the data sources, intermediate data products, and the procedures that were applied to produce the data product. Provenance information is vital in understanding, discovering, validating, and sharing a certain data product as well as the applications and programs used to derive it. In some disciplines such as finance and medicine, it is also mandatory to provide what is called an "audit trail" for audition purpose. In Grids, provenance management has been in general built into a workflow system, from early pioneers such as Chimera [22], to modern scientific workflow systems, such as Swift [53], Kepler [35], and VIEW [34] to support the discovery and reproducibility of scientific results. It has also been built as a standalone service, such as PreServ [29], to facilitate the integration of provenance component in more general computing models, and deal with trust issues in provenance assertion. Using provenance information, scientists can debug workflow execution, validate or invalidate scientific results, and guide future workflow design and data exploration. While provenance has first shown its promise in scientific workflow systems [22] and database systems [47], a long-term vision is that provenance will be useful in other systems as well, necessitating the development of a standard, open, and universal representation and query model. Currently, the provenance challenge series [39] and the open provenance model initiative [38] provide the active forums for these standardization effort and interaction. On the other hand, Clouds are becoming the future playground for e-science research, and provenance management is extremely important in order to track the processes and support the reproducibility of scientific results [45]. Provenance is still an unexplored area in Cloud environments, in which we need to deal with even more challenging issues such as tracking data production across different service providers (with different platform visibility and access policies) and across different software and hardware abstraction layers within one provider. In other words, capturing and managing provenance in Cloud environments may prove to be more difficult than in Grids, since in the latter there are already a few provenance systems and initiatives, however scalable provenance querying [55] and secure access of provenance information are still open problems for both Grids and Clouds environments.

## 2.4 Programming Model

Although programming model in Grid environments does not differ fundamentally from traditional parallel and distributed environments, it is obviously complicated by issues such as multiple administrative domains; large variations in resource heterogeneity, stability and performance; exception handling in highly dynamic (in that resources can join and leave pretty much at any time) environments, etc. Grids primarily target large-scale scientific computations, so it must scale to leverage large number/amount of resources, and we would also naturally want to make programs run fast and efficient in Grid environments, and programs also need to finish correctly, so reliability and fault tolerance must be considered.

We briefly discuss here some general programming models in Grids. MPI (Message Passing Interface) [36] is the most commonly used programming model in parallel computing, in which a set of tasks use their own local memory during computation and communicate by sending and receiving messages. MPICH-G2 [32] is a Grid enabled implementation of MPI. It gives the familiar interface of MPI while providing integration with the Globus Toolkit. Coordination languages also allow a number of possibly heterogeneous components to communicate and interact, offering facilities for the specification, interaction, and dynamic composition of distributed (concurrent) components. For instance, Linda [1] defines a set of coordination primitives to put and retrieve *tuples* from a shared dataspace called the *tuple space*. It has been shown to be straightforward to use such primitives to implement a master-worker parallel scheduler. The Ninf-G GridRPC [37] system integrates a Grid RPC (Remote Procedure Call) layer on top of the Globus toolkit. It publishes interfaces and function libraries in MDS, and utilizes GRAM to invoke remote executables. In Grids, however, many applications are loosely coupled in that the output of one may be passed as input to one or more others—for example, as a file, or via a Web Services invocation. While such "loosely coupled" computations can involve large amounts of computation and communication, the concerns of the programmer tend to be different from those in traditional high performance computing, being focused on management issues relating to the large numbers of datasets and tasks rather than the optimization of interprocessor communication. In such cases, workflow systems [54] suit better in the specification and execution of such applications. More specifically, a workflow system allows the composition of individual (single step) components into a complex dependency graph, and it governs the flow of data and/or control through these components. An example is the Swift system [53], which bridges scientific workflows with parallel computing. It is a parallel programming tool for rapid and reliable specification, execution, and management of large-scale science and engineering workflows. The Swift runtime system relies on the CoG Karajan [33] workflow engine for efficient scheduling and load balancing, and it integrates the Falkon light-weight task execution service for optimized task throughput and resource efficiency [40]. WSRF (Web Services Resource Framework) has emerged from OGSA (Open Grid Service Architecture) [11] as more and more Grid applications are developed as services. WSRF allows Web Services to become stateful, and it provides a set of operations to set and retrieve the states (resources). The Globus Toolkit version 4 contains Java and C implementations of WSRF, most of the Globus core services have been re-engineered to build around WSRF, these altogether will enable service oriented Grid programming model.

MapReduce [13] is only yet another parallel programming model, providing a programming model and runtime system for the processing of large datasets, and it is based on a simple model with just two key functions: "map" and "reduce," borrowed from functional languages. The map function applies a specific operation to each of a set of items, producing a new set of items; a reduce function performs aggregation on a set of items. The MapReduce runtime system automatically partitions input data and schedules the execution of programs in a large cluster of commodity machines. The system is made fault tolerant by checking worker nodes periodically and reassigning failed jobs to other worker nodes. Sawzall is an interpreted language that builds on MapReduce and separates the filtering and aggregation phases for more concise program specification and better parallelization. Hadoop [30] is the open source implementation of the MapReduce model, and Pig is a declarative programming language offered on top of Hadoop. Microsoft has developed the Cosmos distributed storage system and Dryad processing framework, and offers DryadLINQ [31] and Scope as declarative programming model on top of the storage and computing infrastructure. DryadLINQ uses the object oriented LINQ query syntax where Scope provides basic operators similar to those of SQL such as Select, Join, Aggregation etc, both translate the abstract specification into detailed execution plan.

Mesh-up's and scripting (Java Script, PHP, Python etc) have been taking the place of a workflow system in the Cloud world, since there is no easy way to integrate services and applications from various providers. They are essentially data integration approaches, because they take outputs from one service/application, transform them and feed into another. Google App Engine uses a modified Python runtime and chooses Python scripting language for Web application development, the interface to its underlying BigTable storage system is some proprietary query language (named, as you would think, GQL) that is reminiscent of SQL, although all these will probably change. Clouds (such as Amazon Web Services, Microsoft's Azure Services Platform) have generally adopted Web Services APIs where users access, configure and program Cloud services using pre-defined APIs exposed as Web services, and HTTP and SOAP are the common protocols chosen for such services. Although Clouds adopted some common communication protocols such as HTTP and SOAP, the integration and interoperability of all the services and applications remain the biggest challenge, as users need to tap into a federation of Clouds instead of a single Cloud provider.

## 2.5 Application Model

Grids generally support many different kinds of applications, ranging from high performance computing (HPC) to high throughput computing (HTC). HPC applications are efficient at executing tightly coupled parallel jobs within a particular machine with low-latency interconnects and are generally not executed across a wide area network Grid; these applications typically use message passing interface (MPI) to achieve the needed inter-process communication. On the other hand, Grids have also seen great success in the execution of more loosely coupled applications that tend to be managed and executed through workflow systems or other sophisticated and complex applications. Related to HTC applications loosely coupled nature, there are other application classes, such Multiple Program Multiple Data (MPMD), MTC, capacity computing, utility computing, and embarrassingly parallel, each with their own niches [42]. These loosely coupled applications can be composed of many tasks (both independent and dependent tasks) that can be individually scheduled on many different computing resources across multiple administrative boundaries to achieve some larger application goal. Tasks may be small or large, uniprocessor or multiprocessor, compute-intensive or data-intensive. The set of tasks may be static or dynamic, homogeneous or heterogeneous, loosely or tightly coupled. The aggregate number of tasks, quantity of computing, and volumes of data could be small but also extremely large.

On the other hand, Cloud Computing could in principle cater to a similar set of applications. The one exception that will likely be hard to achieve in Cloud Computing (but has had much success in Grids) are HPC applications that require fast and low latency network interconnects for efficient scaling to many processors. As Cloud Computing is still in its infancy, the applications that will run on Clouds are not well defined, but we can certainly characterize them to be loosely coupled, transaction oriented (small tasks in the order of milliseconds to seconds), and likely to be interactive (as opposed to batch-scheduled as they are currently in Grids).

Another emerging class of applications in Grids is scientific gateways [51], which are front-ends to a variety of applications that can be anything from loosely-coupled to tightly-coupled. *A Science Gateway is a community-developed set of tools, applications, and data collections that are integrated via a portal or a suite of applications.* Gateways provide access to a variety of capabilities including workflows, visualization, resource discovery and job execution services through a browser-based user interface (which can arguably hide much of the complexities). Scientific gateways are beginning to adopt a wide variety of Web 2.0 technologies, but to date, much of the developments in Grids and Web 2.0 have been made in parallel with little interaction between them. These new technologies are important enhancements to the ways gateways interact with services and provide rich user interactivity. Although scientific gateways have only emerged in Grids recently, Clouds seem to have adopted the use of gateways to Cloud resources almost exclusively for end-user interaction. The browser and Web 2.0 technologies will undoubtedly play a central role on how users will interact with Grids and Clouds in the future.

## 2.6 Security Model

Clouds mostly comprise dedicated data centers belonging to the same organization, and within each data center, hardware and software configurations, and supporting platforms are in general more homogeneous as compared with those in Grid environments. Interoperability can become a serious issue for cross-data center, cross-administration domain interactions, imagine running your accounting service in Amazon EC2 while your other business operations on Google infrastructure. Grids however build on the assumption that resources are heterogeneous and dynamic, and each Grid site may have its own administration domain and operation autonomy. Thus, security has been engineered in the fundamental Grid infrastructure. The key issues considered are: single sign-on, so that users can log on only once and have access to multiple Grid sites, this will also facilitate accounting and auditing; delegation, so that a program can be authorized to access resources on a user's behalf and it can further delegate to other programs; privacy, integrity and segregation, resources belonging to one user cannot be accessed by unauthorized users, and cannot be tampered during transfer; coordinated resource allocation, reservation, and sharing, taking into consideration of both global and local resource usage policies. The public-key based GSI (Grid Security Infrastructure) protocols are used for authentication, communication protection, and authorization. Furthermore, CAS (Community Authorization Service) is designed for advanced resource authorization within and across communities. Gruber (A Grid Resource Usage SLA Broker) [14] is an example that has distributed policy enforcement points to enforce both local usage policies and global SLAs (Service Level Agreement), which allows resources at individual sites to be efficiently shared in multi-site, multi-VO environments.

Currently, the security model for Clouds seems to be relatively simpler and less secure than the security model adopted by Grids. Cloud infrastructure typically rely on Web forms (over SSL) to create and manage account information for end-users, and allows users to reset their passwords and receive new passwords via Emails in an unsafe and unencrypted communication. Note that new users could use Clouds relatively easily and almost instantly, with a credit card and/or email address. To contrast this, Grids are stricter about its security. For example, although web forms are used to manage user accounts, sensitive information about new accounts and passwords requires also a person to person conversation to verify the person, perhaps verification from a sponsoring person who already has an account, and passwords will only be faxed or mailed, but under no circumstance will they be emailed. The Grid approach to security might be more time consuming, but it adds an extra level of security to help prevent unauthorized access.

Security is one of the largest concerns for the adoption of Cloud Computing. We outline seven risks a Cloud user should raise with vendors before committing [6]: 1) *Privileged user access:* sensitive data processed outside the enterprise needs the assurance that they are only accessible and propagated to privileged users; 2) *Regulatory compliance:* a customer needs to verify if a Cloud provider has external audits and security certifications and if their infrastructure complies with some regulatory security requirements; 3) *Data location:* since a customer will not know where her data will be stored, it is important that the Cloud provider commit to storing and processing data in specific jurisdictions and to obey local privacy requirements on behalf of the customer; 4) *Data segregation:* one needs to ensure that one customer's data is fully segregated from another customer's data; 5) *Recovery:* it is important that the Cloud provider has an efficient replication and recovery mechanism to restore data if a disaster occurs; 6) *Investigative support:* Cloud services are especially difficult to investigate, if this is important for a customer, then such support needs to be ensured with a contractual commitment; and 7) *Long-term viability:* your data should be viable even the Cloud provider is acquired by another company.

## 3 Conclusions and lights to the future

In this paper, we show that Clouds and Grids share a lot commonality in their vision, architecture and technology, but they also differ in various aspects such as security, programming model, business model, compute model, data model, applications, and abstractions. We also identify challenges and opportunities in both fields. We believe a close comparison such as this can help the two communities understand, share and evolve infrastructure and technology within and across, and accelerate Cloud Computing from early prototypes to production systems.

What does the future hold? We will hazard a few predictions, based on our beliefs that the economics of computing will look more and more like those of energy. Neither the energy nor the computing grids of tomorrow will look like yesterday's electric power grid. Both will move towards a mix of micro-production and large utilities, with increasing numbers of small-scale producers (wind, solar, biomass, etc., for energy; for computing, local clusters and embedded processors—in shoes and walls) co-existing with large-scale regional producers, and load being distributed among them dynamically. Yes, computing isn't really like electricity, but we do believe that we will nevertheless see parallel evolution, driven by similar forces.

In building this distributed "Cloud" or "Grid", we will need to support on-demand provisioning and configuration of integrated "virtual systems" providing the precise capabilities needed by an end-user. We will need to define protocols that allow users and service providers to discover and hand off demands to other providers, to monitor and manage their reservations, and arrange payment. We will need tools for managing both the underlying resources and the resulting distributed computations. We will need the centralized scale of today's Cloud utilities, and the distribution and interoperability of today's Grid facilities.

Unfortunately, at least to date, the methods used to achieve these goals in today's commercial clouds have not been open and general purpose, but instead been mostly proprietary and specialized for the specific internal uses (e.g., large-scale data analysis) of the companies that developed them. The idea that we might want to enable interoperability between providers (as in the electric power grid) has not yet surfaced. Grid technologies and protocols speak precisely to these issues, and should be considered.

Some of the required protocols and tools will come from the smart people from the industry at Amazon, Google, Yahoo, Microsoft, and IBM. Others will come from the smart people from academia and government labs. Others will come from those creating whatever we call this stuff after Grid and Cloud. It will be interesting to see to what extent these different communities manage to find common cause, or instead proceed along parallel paths.